\documentclass{article}

\usepackage{spconf,amsmath,graphicx}
\usepackage{graphicx,algorithm,algorithmicx,bm,amsmath,amssymb,rotating,graphicx,cite,amsthm,calc,color,epstopdf,xcolor,mathrsfs,dsfont}
\usepackage{lettrine}
\usepackage{cite}
 \usepackage{amsfonts}
\usepackage{longtable}
\usepackage{float}
\usepackage{subcaption}
\usepackage{flushend}
\usepackage{multirow}
\definecolor{icassp}{rgb}{0.21,0.49,0.74}
\usepackage[pagebackref,breaklinks,colorlinks,citecolor=icassp]{hyperref}
\usepackage{cleveref}
\usepackage{booktabs}

\begin{document}
\title{Self-Prompting Polyp Segmentation in Colonoscopy Using Hybrid YOLO-SAM 2 Model}

\name{Mobina Mansoori$^\dag$, Sajjad Shahabodini$^\dag$, Jamshid Abouei $^{\dag\dag}$,\vspace{-.14in}}
\address{\textit{Konstantinos N. Plataniotis$^\ddag$, and Arash Mohammadi$^\dag$\thanks{This work was partially supported by the Natural Sciences and Engineering Research Council (NSERC) of Canada through the NSERC Discovery Grant RGPIN-2023-05654}}\\\\
$~^\dag$ Intelligent Signal \& Information Processing (I-SIP) Lab, Concodia University, Canada \\$~^\ddag$ Edward S. Rogers Sr. Department of Electrical and Computer Engineering, University of Toronto \\$~^{\dag\dag}$ Department of Electrical Engineering, Yazd University, Iran}

\ninept
\maketitle
\begin{abstract}
Early diagnosis and treatment of polyps during colonoscopy are essential for reducing the incidence and mortality of Colorectal Cancer (CRC). However, the variability in polyp characteristics and the presence of artifacts in colonoscopy images and videos pose significant challenges for accurate and efficient polyp detection and segmentation. This paper presents a novel approach to polyp segmentation by integrating the Segment Anything Model (SAM 2) with the YOLOv8 model. Our method leverages YOLOv8’s bounding box predictions to autonomously generate input prompts for SAM 2, thereby reducing the need for manual annotations. We conducted exhaustive tests on five benchmark colonoscopy image datasets and two colonoscopy video datasets, demonstrating that our method exceeds state-of-the-art models in both image and video segmentation tasks. Notably, our approach achieves high segmentation accuracy using only bounding box annotations, significantly reducing annotation time and effort. This advancement holds promise for enhancing the efficiency and scalability of polyp detection in clinical settings \url{https://github.com/sajjad-sh33/YOLO_SAM2}.
\end{abstract}

\section{Introduction}
\begin{figure*}[h]
\centering
\includegraphics[width=\textwidth]{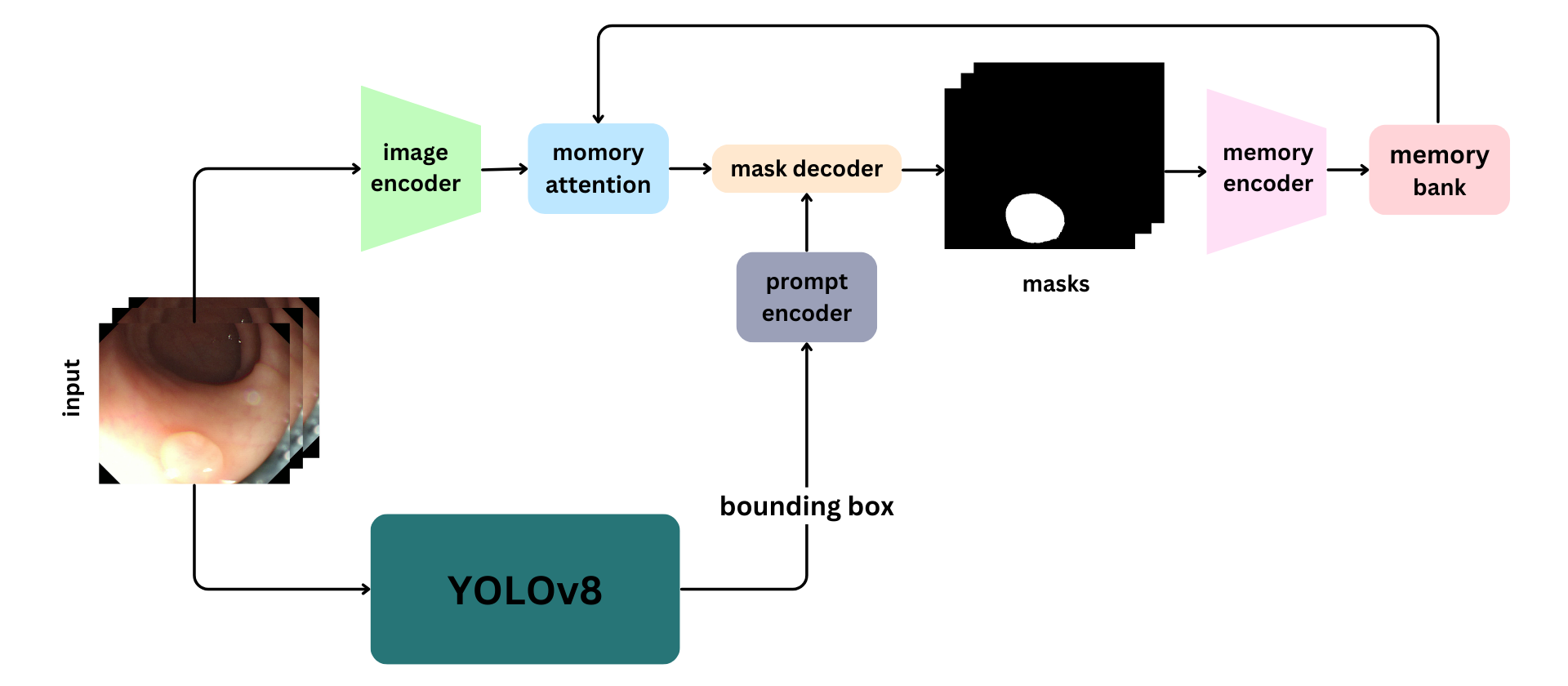}
\caption{Architecture of the Self-Prompting YOLO-SAM 2 Model for Polyp Segmentation.}
\label{fig1}
\end{figure*}
Colorectal Cancer (CRC) is one of the leading causes of cancer-related deaths worldwide, with millions of new cases diagnosed each year \cite{sung2021global}. Early detection and removal of polyps during colonoscopy significantly reduces the incidence and mortality of CRC. However, the accuracy and efficiency of polyp detection and segmentation remain challenging due to the variability in polyp size, shape, and appearance, as well as the presence of various artifacts and noise in colonoscopy images and videos.

Computer-Aided Diagnosis (CAD) systems for colonoscopy have shown significant potential in enhancing annotation efficiency and reducing the time required for diagnosis. The advent of deep learning has led to the development of numerous neural networks tailored for medical image segmentation \cite{chen2024efficient}. Most current polyp segmentation algorithms rely on variations of the UNet \cite{ronneberger2015u} architecture to segment polyps \cite{yin2023semantic}. Additionally, some approaches utilize Res2Net \cite{gao2019res2net} as the backbone and incorporate various regularization techniques to improve segmentation accuracy \cite{zhou2023cross}. Recently, attention-based methods have also been introduced to further enhance performance \cite{wei2021shallow}.

Despite these advancements, the process of annotating medical images remains labor-intensive and costly, as it typically requires medical expertise. This challenge has sparked interest in transfer learning, which applies knowledge from large-scale natural image datasets to specific medical imaging tasks. Recent breakthroughs in foundational models, particularly the Segment Anything Model (SAM) \cite{kirillov2023segment}, have shown exceptional performance in generating high-quality object masks from various input prompts. SAM’s success in multiple computer vision benchmarks has gained significant attention for its application in medical image segmentation \cite{ma2024segment}, including polyp image segmentation \cite{li2024polyp, rahman2024pp}. The introduction of SAM 2 \cite{ravi2024sam} has further improved real-time segmentation, enabling the processing of entire video sequences based on single-frame annotations. This advancement reduces user interaction time and enhances overall performance, making SAM 2 a valuable tool in medical diagnostics and other applications \cite{zhang2024unleashing, mansoori2024polyp}.

Despite the strong zero-shot capabilities of SAM 2 for segmenting medical images, it typically requires input prompts provided by human experts. This reliance on manual input limits the efficiency and scalability of the segmentation process. To address this limitation, we propose a self-prompting segmentation model that autonomously generates input prompts by integrating the YOLOv8 \cite{jocher2023yolo} model's pre-trained capabilities. By combining YOLO’s bounding box predictions with SAM 2’s segmentation capabilities, our method aims to enhance the accuracy and efficiency of polyp segmentation in colonoscopy images and videos. In this work, we utilize only bounding box data for training our model to perform the segmentation task. This approach significantly reduces annotation time compared to previous methods that required detailed ground truth segmentation masks for training. By leveraging bounding box annotations, our model can achieve high segmentation accuracy with less manual effort, making it more practical for large-scale applications. This approach not only tackles variability in polyp features but also reduces the computational load for large-scale segmentation tasks.

\section{Methodology}
Our proposed model integrates two state-of-the-art algorithms, YOLO and SAM 2, to effectively detect and segment polyps in colonoscopy images and videos. The process begins with the YOLO model, which identifies potential polyps and places bounding boxes around them. These bounding boxes are then employed as input prompts for the SAM 2 model, which performs precise segmentation of the polyps based on the provided coordinates, Fig.~\ref{fig1}.

A pre-trained YOLOv8 model is employed for its superior speed and accuracy in real-time object detection tasks. YOLOv8 processes the colonoscopy images through a Convolutional Neural Network (CNN), extracting essential features and predicting bounding boxes around potential polyps. The box coordinates of these bounding boxes are then passed to the SAM 2 model as prompts.

The SAM 2 model, known for its lightweight nature and high accuracy, refines the detection results provided by YOLOv8. It uses bounding box coordinates to perform detailed segmentation, delineating the exact boundaries of the polyps. The SAM 2 model architecture consists of several key components: \textbf{Image Encoder}: Utilizes a transformer-based architecture to extract high-level features from both images and video frames. This component is responsible for understanding visual content at each timestep. \textbf{Prompt Encoder}: Processes user-provided prompts to guide the segmentation task. This allows SAM 2 to adapt to user input and target specific objects within a scene. \textbf{Memory Mechanism}: Includes a memory encoder, memory bank, and memory attention module. These components collectively store and utilize information from past frames, enabling the model to maintain consistent object tracking over time. \textbf{Mask Decoder}: Produces the final segmentation masks based on the encoded image features and prompts.

Due to the lightweight nature of both models, it is still possible to use their combination as a real-time segmentation model to segment polyp videos. During training, only the YOLOv8 model would be fine-tuned based on the bounding box dataset, while the SAM 2 weights are frozen and not fine-tuned.

It is important to emphasize that in this study, the YOLOv8 model is dedicated solely to generating bounding boxes rather than segmentation masks. Our approach focuses on utilizing only bounding box data to train the overall segmentation model, leveraging the zero-shot capabilities of SAM 2 to minimize the need for extensive data annotation. Additionally, our findings indicate that SAM 2 provides superior segmentation performance than YOLOv8.
\begin{table*}[t!]
\caption{A quantitative comparison of five benchmark datasets with state-of-the-art (SOTA) methods is presented. The best performance is highlighted in bold.}
\fontsize{11pt}{9pt}\selectfont
\renewcommand{\arraystretch}{1.5}
\begin{center}
\begin{tabular}{@{}c|cc|cc|cc|cc|cc@{}}
\hline
\multirow{2}{*}{\textbf{Methods}}& \multicolumn{2}{c|}{\textbf{CVC-ClinicDB}} & \multicolumn{2}{c|}{\textbf{Kvasir-SEG}} & \multicolumn{2}{c|}{\textbf{CVC-ColonDB}} & \multicolumn{2}{c|}{\textbf{ETIS}} & \multicolumn{2}{c}{\textbf{CVC-300}} \\
\cline{2-11}
                         & \textbf{\textit{mIoU}}           & \textbf{\textit{mDice}}           & \textbf{\textit{mIoU}}        & \textbf{\textit{mDice}}          & \textbf{\textit{mIoU}}          & \textbf{\textit{mDice}}             & \textbf{\textit{mIoU}}       &\textbf{\textit{mDice}}         & \textbf{\textit{mIoU}}         & \textbf{\textit{mDice}}            \\
 \cline{1-11} 
UNet\cite{ronneberger2015u}                     & 0.755           & 0.823          & 0.746        & 0.818       & 0.436          & 0.504          & 0.335       & 0.398      & 0.627         & 0.710         \\

UNet++\cite{zhou2018unet++}                   & 0.729           & 0.794          & 0.744        & 0.821       & 0.408          & 0.482          & 0.344       & 0.401      & 0.624         & 0.707         \\

MSEG   \cite{huang2021hardnet}                  & 0.864           & 0.909          & 0.839        & 0.897       & 0.666          & 0.735          & 0.630       & 0.700      & 0.804         & 0.874         \\
SANet     \cite{wei2021shallow}               & 0.859           & 0.916          & 0.847        & 0.904       & 0.669          & 0.752          & 0.654       & 0.750      & 0.815         & 0.888         \\
MSNet   \cite{zhao2021automatic}                 & 0.869           & 0.918          & 0.849        & 0.905       & 0.668          & 0.747          & 0.650       & 0.720      & 0.796         & 0.862         \\
SSFormer        \cite{wang2022stepwise}         & 0.855           & 0.906          & 0.864        & 0.917       & 0.721          & 0.802          & 0.720       & 0.796      & 0.827         & 0.895         \\
CFA-Net      \cite{zhou2023cross}           & 0.883           & 0.933          & 0.861        & 0.915       & 0.665          & 0.743          & 0.655       & 0.732      & 0.827         & 0.893         \\
Polyp-PVT    \cite{dong2021polyp}            & 0.905           & 0.948          & 0.864        & 0.917       & 0.727          & 0.808          & 0.706       & 0.787      & 0.833         & 0.900         \\
SAM-Path    \cite{zhang2023sam}              & 0.644           & 0.750          & 0.730        & 0.828       & 0.516          & 0.632          & 0.442       & 0.555      & 0.756         & 0.844         \\
SurgicalSAM \cite{yue2024surgicalsam}             & 0.505           & 0.644          & 0.597        & 0.740       & 0.330          & 0.460          & 0.238       & 0.342      & 0.472         & 0.623         \\
 IC-PolypSeg \cite{ chen2024efficient}            & 0.89           & 0.938           & 0.859         & 0.91        & 0.729              & 0.807              & 0.692           & 0.774          & {{ 0.844}}             &{{0.909}}               \\
FAGF-Net \cite{li2024frequency}                &0.898& 0.943        &{\textbf{0.879 }}         & {\textbf{0.927}}        & 0.738          & 0.820          & 0.724       & 0.801      & 0.837         & 0.903         \\
\cline{1-11} 
Yolo-SAM         &0.810      &0.895       & 0.742    &   0.852      &0.808        &0.893          & 0.875      &0.933      & {\textbf{0.865}}         &{\textbf{0.925}}          \\
\hline
\cline{1-11} 
\textbf{Yolo-SAM 2}        & {\textbf{0.909}}           & {\textbf{0.951}}         & {{0.764 }}         &{{ 0.866}}         & {\textbf{0.848}}           & {\textbf{0.918}}          &{\textbf{0.904}}       & {\textbf{0.949}}      &{{0.817}}            & {{0.889}}        \\
\hline
\end{tabular}
\label{tab1}
\end{center}
\end{table*}
\section{Experiments and Results}
\subsection{Datasets}
In order to assess the performance of the SAM 2 model, we carried out comparative experiments utilizing five well-known benchmark colonoscopy image datasets, along with two additional video colonoscopy datasets. The following sections provide detailed descriptions of each dataset. 1) \textbf{Kvasir-SEG} \cite{jha2020kvasir}: Curated by the Vestre Viken Health Trust in Norway, this dataset includes 1,000 polyp images and their corresponding ground truth from colonoscopy video sequences. 2) \textbf{CVC-ClinicDB} \cite{tajbakhsh2015automated}: Created in collaboration with the Hospital Clinic of Barcelona, Spain, it contains 612 images from colonoscopy examination videos, originating from 29 different sequences. 3) \textbf{CVC-ColonDB} \cite{tajbakhsh2015automated}: This dataset comprises 380 polyp images, each with its corresponding ground truth, captured at a resolution of 500 × 570 pixels from 15 distinct videos. 4) \textbf{ETIS} \cite{silva2014toward}: It includes 196 polyp images, each captured at a resolution of 966 × 1225 pixels, aiding research in polyp detection and analysis. 5) \textbf{CVC-300} \cite{vazquez2017benchmark}: Comprising 60 polyp images, each captured at a resolution of 500 × 574 pixels. 6) \textbf{PolypGen} \cite{ali2023multi}: A comprehensive dataset for polyp detection and segmentation, including 1,537 polyp images, 4,275 negative frames, and 2,225 positive video sequences, collected from six medical centers across Europe and Africa. 7) \textbf{SUN-SEG} \cite{ji2022video, misawa2021development}: This dataset includes 158,690 frames from 113 colonoscopy videos, with detailed annotations for each frame.

\subsection{Implementation Details}
As mentioned earlier, the SAM model remains frozen, and only the YOLOv8 model is trained. After conducting experiments, we chose the YOLOv8 medium version due to its superior performance compared to the large and small versions. Despite having 25 million parameters, it effectively handles real-time video segmentation when combined with the SAM 2 model. Additionally, we use the SAM 2 large model, which contains 224.4 million parameters. Despite its larger size, it maintains real-time performance, processing approximately 44 frames per second. This makes it suitable for applications requiring high accuracy and speed, such as video analysis and interactive segmentation tasks. For training, 80\% of the dataset is used, with the remaining 20\% reserved for evaluation. We set the batch size to 64 and the input image size to 680. We implemented our model using A100 (40 GB) GPU.
\subsection{Evaluation Metrics}
To assess the performance of our model, we use the following metrics. \textbf{Intersection over Union (IoU)}: Measures the overlap between the predicted segmentation mask and the ground truth mask. \textbf{Dice Coefficient}: Evaluates the similarity between the predicted and ground truth masks. Additionally, for video datasets, we employ four other metrics, including F-measure ($F_{\beta}^{\text{mn}}$), sensitivity (Sen), enhanced-alignment measure ($E_{\phi}^{\text{mn}}$), and structure-measure ($S_{\alpha}$).
\subsection{Comparison with State-of-the-art Methods}
In this section, we evaluate our model’s performance against several state-of-the-art methods for polyp segmentation in images. We provide both quantitative metrics and qualitative visualizations to highlight the strengths of our approach.

 \Cref{tab1} presents a quantitative comparison of our proposed model with various state-of-the-art methods across five publicly available polyp segmentation datasets, as mentioned in Section 3.1, using the metrics discussed in Section 3.3 Specifically, we compared our model with several CNN and ViT models, as well as recent SAM-based segmentation techniques. In our study, CNN-based models include UNet \cite{ronneberger2015u}, UNet++ \cite{zhou2018unet++}, MSEG \cite{huang2021hardnet}, SANet \cite{wei2021shallow}, MSNet \cite{zhao2021automatic},  and CFA-Net \cite{zhou2023cross}. For transformer-based models, we evaluated SSFormer \cite{wang2022stepwise} and Polyp PVT \cite{dong2021polyp}. Furthermore, we explored the impact and effectiveness of SAM-Path  \cite{zhang2023sam}, SurgicalSAM  \cite{yue2024surgicalsam}, IC-PolypSeg \cite{chen2024efficient} and FAGF-Net \cite{li2024frequency} of which are built upon the SAM model.\\
Based on the results, we can conclude that YOLO-SAM 2 is capable of effectively locating and segmenting polyps without additional training. More importantly, among all image segmentation methods, YOLO-SAM 2 has achieved the highest performance across some scores by a considerable margin (e.g., 9.8\%, 11\% in mDice, mIoU on CVC-ColonDB  \cite{tajbakhsh2015automated}, 14.8\%, 18\% in mDice, mIoU on ETIS-LaribPolypD \cite{silva2014toward} than the previous-best methods).

\begin{figure}[t!]
  \centering
 \includegraphics[width=\columnwidth]{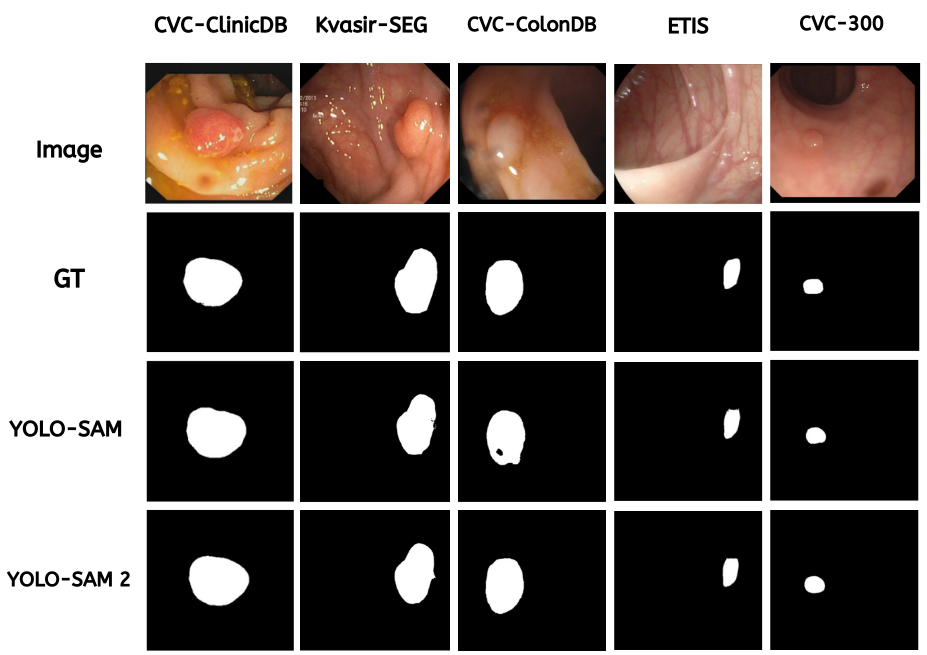}
  \caption{Qualitative Assessment of five polyp segmentation datasets using YOLO-SAM and YOLO-SAM 2.}
  \label{Fig2}
  \vspace{-20pt}
\end{figure}

The qualitative assessment of YOLO-SAM 2 for polyp segmentation is also explored. \cref{Fig2} showcases the visualization outcomes compared to the YOLO-SAM model, utilizing two chosen benchmark datasets. Remarkably, the YOLO-SAM 2 model exhibits enhanced performance, producing segmentation results that are very close to the actual ground truth.
\begin{table*}[t!]
\caption{Quantitative comparison of the four subsets of the SUN-SEG dataset, highlighting the top performance in bold.}
\centering
\setlength\tabcolsep{9.7pt}
\begin{tabular}{c| c| c c c c c| c c c c c}
    \toprule
    \multirow{2}{*}{\textbf{}} & \multirow{2}{*}{\textbf{Methods}} & \multicolumn{5}{c}{\textbf{SUN-SEG-Easy}} & \multicolumn{5}{c}{\textbf{SUN-SEG-Hard}} \\
    \cmidrule(lr){3-7} \cmidrule(lr){8-12}
    & & {\textit{$S_{\alpha}$}} & {\textit{$E_{\phi}^{\text{mn}}$}} & {\textit{$F_{\beta}^{\text{mn}}$}} &{\textit{Sen}} & {\textit{Dice}} & {\textit{$S_{\alpha}$}} & {\textit{$E_{\phi}^{\text{mn}}$}} & {\textit{$F_{\beta}^{\text{mn}}$}} & {\textit{Sen}} & {\textit{Dice}} \\
    \midrule
    \multirow{7}{*}{\rotatebox{90}{\textbf{Seen}}}&2/3D \cite{puyal2020endoscopic}& 0.895& 0.909& 0.853& 0.808& 0.856& 0.849& 0.868& 0.805& 0.726& 0.809 \\
&PNS-Net \cite{ji2021progressively} &0.906 &0.910 &0.860 &0.827 &0.861 &0.870 &0.892 &0.822 &0.774 &0.823\\
&PNS+ \cite{ji2022video} &0.917 &0.924 &0.878 &0.837 &0.888 &0.887 &0.929 &0.849 &0.780 &0.855\\
&FLA-Net \cite{lin2023shifting} &0.906 &0.922 &0.867 &0.851 &0.875 &0.859 &0.892 &0.810 &0.785 &0.809\\
&SLT-Net \cite{cheng2022implicit} &0.927 &0.961 &0.914 &0.888 &0.906 &0.894 &0.943 &0.874 &0.851 &0.866\\

\cline{2-12}
&\textbf{YOLO-SAM 2} &\textbf{0.937}&\textbf{0.971} &\textbf{0.958} & \textbf{0.923}& \textbf{0.945} & \textbf{0.893}& \textbf{0.931}&\textbf{0.902}&\textbf{0.805}&\textbf{0.865}\\
   \midrule

    \multirow{7}{*}{\rotatebox{90}{\textbf{Unseen}}}&2/3D \cite{puyal2020endoscopic} &0.786 &0.777 &0.708 &0.603 &0.722 &0.786 &0.775 &0.688 &0.607 &0.706\\
&PNS-Net \cite{ji2021progressively} &0.767 &0.744 &0.664 &0.574 &0.676 &0.767 &0.755 &0.656 &0.579 &0.675\\
&PNS+\cite{ji2022video} &0.806 &0.798 &0.730 &0.630 &0.756 &0.797 &0.793 &0.709 &0.623 &0.737\\
&FLA-Net\cite{lin2023shifting} &0.722 &0.697 &0.597 &0.506 &0.636 &0.721 &0.701 &0.592 &0.522 &0.628\\
&SLT-Net \cite{cheng2022implicit}] &0.848 &0.893 &0.817 &0.747 &0.792 &0.844 &0.904 &0.795 &0.760 &0.781\\
\cline{2-12} 
&\textbf{YOLO-SAM 2}& \textbf{0.90} & \textbf{0.938} & \textbf{0.938} & \textbf{0.837} & \textbf{0.90} & \textbf{0.894} & \textbf{0.941} & \textbf{0.932} & \textbf{0.852} & \textbf{0.902}\\
    \bottomrule
\end{tabular}
    \label{tab3}
\end{table*}
\begin{table}[h!]
\caption{A quantitative comparison of 23 sequence videos of the Polypgen dataset with state-of-the-art methods is presented. The best performance is highlighted in bold.}
\begin{center}
\setlength\tabcolsep{4.5pt}
\begin{tabular}{@{}c|ccccc@{}}
\hline
\textbf{Methods} & \textbf{\textit{mDice}} & \textbf{\textit{mIoU}} & \textbf{\textit{Precision}} & \textbf{\textit{Recall}} & \textbf{\textit{F2}} \\
\hline
UNet\cite{ronneberger2015u} & 0.4559 & 0.4049 & 0.5762 & 0.6307 & 0.4668 \\
UNet++\cite{zhou2018unet++} & 0.4772 & 0.4272 & 0.6269 & 0.6198 & 0.4876 \\
ResU-Net++\cite{jha2019resunet++} & 0.2105 & 0.1589 & 0.2447 & 0.5095 & 0.2303 \\
MSEG\cite{huang2021hardnet} & 0.4662 & 0.4171 & 0.6120 & 0.6217 & 0.4757 \\
ColonSegNet\cite{jha2021real} & 0.3574 & 0.3058 & 0.4804 & 0.5296 & 0.3533 \\
UACANet\cite{kim2021uacanet} & 0.4748 & 0.4155 & 0.6108 & 0.6357 & 0.4886 \\
UNeXt\cite{valanarasu2022unext} & 0.2998 & 0.2457 & 0.3661 & 0.5658 & 0.3201 \\
TransNetR\cite{jha2024transnetr} & 0.5168 & 0.4717 & 0.7881 & 0.5777 & 0.5105 \\
\cline{1-6} 
\textbf{YOlO-SAM 2} & \textbf{0.808} & \textbf{0.678} &\textbf{0.858} &\textbf{0.764} & \textbf{0.781}\\
\hline
\end{tabular}
\label{tab2}
\vspace{-25pt}
\end{center}
\end{table}
\subsection{Quantitative Results on Video Polyp Segmentation}
In this section, we assess the performance of our proposed model for polyp video segmentation using the SUN-SEG and PolypGen datasets. \Cref{tab3} presents the results for four sub-test sets: SUN-SEG-Seen-Hard, SUN-SEG-Seen-Easy, SUN-SEG-Unseen-Hard, and SUN-SEG-Unseen-Easy. The terms Easy/Hard refer to the difficulty levels of the samples to be segmented, Seen indicates that the clips are from the same video as the training set but do not overlap, and Unseen indicates that the clips are from videos that do not overlap with the training set. YOLO-SAM 2 has outperformed the previous best method by achieving a 7.5\% higher dice score for SUN-SEG-Unseen-Easy and an 8\% higher dice score for SUN-SEG-Unseen-Hard. Notably, these improvements were achieved without training the model with segmentation masks; instead, only bounding box annotations were used for training. This approach distinguishes our method from previous works. Additionally, the results for the PolypGen dataset, as shown in \cref{tab2}, demonstrate that the YOLO-SAM 2 model significantly improves video segmentation performance, achieving a remarkable 20.7\% increase in mean intersection over union (mIoU) compared to previous state-of-the-art methods.
\section{Conclusion}
In this paper, we introduced a self-prompting segmentation model that combines the strengths of SAM 2 and YOLOv8 for real-time polyp detection in colonoscopy images and videos. Our approach addresses the limitations of manual input prompts by leveraging YOLOv8’s pre-trained capabilities to generate bounding box predictions, which are then used by SAM 2 for accurate segmentation. Through comprehensive experiments on multiple benchmark datasets, we demonstrated that our model achieves superior performance compared to existing state-of-the-art methods. The significant improvements in segmentation accuracy, coupled with the reduced need for detailed ground truth masks, highlight the practicality and efficiency of our method for large-scale applications. Future work will focus on further optimizing the model for real-time clinical deployment and exploring its potential for other medical imaging tasks.


\bibliographystyle{IEEEtran}
\footnotesize
\bibliography{refs}

\end{document}